# Bad-Metallic Behavior of Doped Mott Insulators

Jure Kokalj

*J. Stefan Institute, SI-1000 Ljubljana, Slovenia and*
*Faculty of Civil and Geodetic Engineering, University of Ljubljana, SI-1000 Ljubljana, Slovenia*

Employing Nernst-Einstein decomposition $\sigma = e^2 \chi_c D$ of the conductivity $\sigma$ onto charge susceptibility (compressibility) $\chi_c$ and diffusion constant $D$, we argue that the bad-metallic behavior of $\sigma$ in the regime of high temperatures and lightly doped insulator is dominated by the strong temperature and doping dependence of $\chi_c$. In particular, we show how at small dopings $\chi_c$ strongly decreases towards undoped-insulating values with increasing temperature and discuss simple picture leading to the linear-in-temperature resistivity with the prefactor increasing inversely with decreasing concentration ($p$) of doped holes, $\rho \propto T/p$. On the other hand, $D$ shows weak temperature and doping dependence in the corresponding regime. We support our arguments by numerical results on the two dimensional Hubbard model and discuss the proposed picture from the experimental point of view.

Many materials show puzzling metallic behavior in which the resistivity is linear-in-temperature up to a very high temperature, e.g. 1000 K, and smoothly crosses the Mott-Ioffe-Regel (MIR) limit without any indication [1]. Such behavior with resistivity being metallic (increasing with temperature) and larger than MIR limit is termed the bad-metallic behavior, which is frequently associated also with the breakdown of the Fermi-liquid concept and the loss of coherent quasiparticles. Understanding this behavior is one of the central challenges of the solid state physics ever since its discovery [2] in the cuprate high-temperature superconductors, since it might be the key towards understanding of the superconductivity. Bad-metallic behavior is observed also in may other classes of materials like pnictides [3–5], fullerenes [6], vanadium dioxide [7], ruthenates [8], organic charge transfer salts [9] and nickelates [10].

Despite this ubiquitous behavior, consensus on its proper understanding is still missing. In 1970s it has been discussed in terms or electron-phonon scattering [11, 12], followed by discussion of strong electron-electron interactions and correlations [1, 13, 14], while more recent proposals include quantum criticality [15–17] and a bound on diffusion constant motivated by holographic duality [18]. It has been discussed also with approaches aiming at lower temperatures, e.g., marginal Fermi liquid [19], spin fluctuation [20, 21] and charge density wave [22] scenarios.

We use the Nernst-Einstein relation

$$\sigma = e^2 \chi_c D, \quad (1)$$

which relates conductivity $\sigma$ to the charge susceptibility (compressibility) $\chi_c = \partial n/\partial \mu$ and diffusion constant $D$. $e$ is electronic charge, $n$ is electronic density and $\mu$ is chemical potential. We argue that in doped insulators at high temperatures ($T$) the linear-in-$T$ resistivity arrises due to strong $T$ dependence of $\chi_c \propto 1/T$ and is therefore a static effect, while the diffusion constant $D$ has a rather weak $T$ dependence. We also show, that at small dopings ($p$) $\chi_c$ gets strongly suppressed with increasing $T$ due to approaching insulating values, which strongly increases the resistivity $\rho = 1/\sigma$ and naturally leads to the resistivity crossing the MIR limit. We also show that $\chi_c$ has strong $p$ dependence and discuss possible effects of antiferromagnetic (AFM) correlations.

*Nernst-Einstein relation.* Eq. (1) deconstructs $\sigma$ onto $\chi_c$ and $D$. $\chi_c$ measures the change of electronic density due to changes of $\mu$ and since $\mu$ plays essentially the same role as a static uniform electric potential $e\phi$, $\chi_c$ can be seen as the measure of electronic density redistribution in the presence of slowly varying external electric potential. A particularly interesting situation appears for doped antiferromagnetic Mott-insulators, where in the insulator $\chi_c = 0$, while with doping a transition from an insulator to a metal appears with discontinuous jump or divergence of $n$ vs. $\mu$, i.e. $\chi_c = \infty$. These two extreme limits of $\chi_c$ very close in the phase diagram can lead to strong $T$ and $p$ dependence of $\chi_c$ and in turn of $\rho$, simply due to $T$-broadening effect. On the other hand, if $\chi_c$ measures the tension for charge redistribution in the presence of external potential, $D$ describes the rate at which the electronic density redistributes and is therefore a dynamic quantity, e.g., it can be related to the mean square velocity $\langle v^2 \rangle$ and the scattering rate $1/\tau$, $D = \langle v^2 \rangle \tau/2$. We note that Eq. (1) can be derived on the operator level [23] and its validity is not limited to only certain phases, e.g., to Fermi liquids, and should be valid also in the incoherent bad-metallic regime [24].

*Hubbard model results.* To explore the behavior of $\chi_c$ and $D$ of doped insulators we use the Hubbard model on a two dimensional lattice, which allows the description of the insulating phase (with possibility of AFM ordering) as well as of the metallic doped insulator. The model Hamiltonian is

$$H = -t \sum_{\langle i,j \rangle, s} (c^\dagger_{i,s} c_{j,s} + \text{h.c.}) + U \sum_i n_{i,\uparrow} n_{i,\downarrow} - \mu \sum_{i,s} n_{i,s}. \quad (2)$$

$t$ is the hopping amplitude, $c_{i,s}$ ($c^\dagger_{i,s}$) is the annihilation (creation) operator of electron on a site $i$ with spin $s$ ($\uparrow$ or $\downarrow$), $\langle i,j \rangle$ denotes nearest-neighbor sites on a lattice (we focus only on a square and triangular lattice), $U$ is the onsite Coulomb repulsion and $n_{i,s} = c^\dagger_{i,s} c_{i,s}$. We evaluate quantities for this model by using numerical Finite-temperature Lanczos method (FTLM) [13, 25] on finite clusters (16 sites), which allows a precise evaluation of physical properties at high $T$ and is therefore suitable to tackle these questions. At low $T$ finite-size effects appear [24]. We further set $\hbar = k_B = 1$.

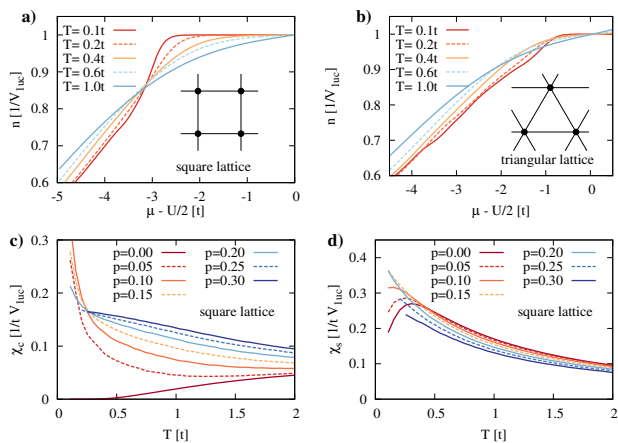

Figure 1. (Color online) Electron density $n$ vs. chemical potential $\mu$ and charge susceptibility ($\chi_c = \partial n/\partial \mu$) show strong $T$ dependence at low dopings of Mott insulator. Results further suggest that $\chi_c$ is at low dopings increased by AFM correlations. a) $n$ vs. $\mu$ for several $T$ and square lattice, showing gapped behavior at $n = 1$ and finite slope at finite dopings ($n < 1$) with particularly large slope at small dopings. b) $n$ vs. $\mu$ for several $T$ and for triangular lattice, showing smaller slope at finite dopings than square lattice. This suggest $\chi_c$ at low dopings can be increased by AFM correlations, which are larger for square lattice than for frustrated triangular one. This is further supported by comparison of panel c), which shows $\chi_c$ vs. $T$ for several doping and its increase at low $T$ and low doping, with panel d), which shows in the same regime suppressed uniform spin susceptibility $\chi_s$. This suppression is due to stronger AFM correlations [26]. Results are calculated for $U = 10t$ and $V_{\text{luc}}$ denotes one-unit-cell volume.

In Fig. 1a we show $n(\mu)$ calculated on a square lattice for several $T$ and for $U = 10t$, which is within an insulating regime for half-filling ($n = 1$). At low $T \sim 0.1t$ a gapped behavior with $\chi_c = \partial n/\partial \mu = 0$ is observed at $n = 1$, while at finite hole dopings $\chi_c$ is finite. With increasing $T$ the dependence of $n$ on $\mu$ is smoothed. For $n = 1$ the behavior of $\chi_c$ is activated obtaining finite values at $T$ comparable to the charge gap $\Delta_c$ (see Fig. 1c and Ref. 25). On the other hand, it is particularly evident that for small dopings ($n = 0.95$, $p = 1 - n = 0.05$) $\chi_c$ gets strongly reduced with increasing $T$ and that $\mu$ is rapidly moved into the gapped regime for low $T$. This strong $T$ dependence of $\chi_c$ is reflected in strong $T$ dependence of $\rho$ as shown below in Fig. 2. Such behavior originates in a simple electron number conservation, which makes the behavior of doped system similar to insulating one with increasing $T$ [24]. We stress that this situation appears for any doped charge-gapped system, as long as the density of electrons is conserved (e.g. semiconductor, band insulator, Mott insulator, etc.).

$\chi_c$ of a lightly doped Mott insulator has additional feature in contrast to, e.g., semiconductor or band insulator. At $T = 0$ and for $U \gg t$ the increase of $U$ for $\delta U$ leads to the increase of $\Delta_c$ at $n = 1$ to roughly $\Delta_c + \delta U$, but due to the same increase of $\mu_{\max}$ at which the band is completely filled ($n = 2$) to $\mu_{\max} + \delta U$, such increase of $U$ does not (at least on average) change $\chi_c$ for finite dopings ($n \ne 1$) [24]. The effect of strong AFM correlations is different. The appearance of AFM correlations decreases the energy of Mott-insulating state for an order of exchange energy $J$, makes it more stable and increases $\Delta_c$ (see, e.g., Ref. 27 for indication of such behavior). For fixed $U$ and in turn fixed $\mu_{\max}$ as well as $\mu_{\min}$, this needs to be accompanied with the increase of $\chi_c$ for $n \ne 1$. Such increase is seen in Fig. 1a and 1c at low dopings (e.g. $p \sim 0.05$) and at low $T$ ($\sim 0.1t$), while it does not appear in a triangular lattice as shown in Fig. 1b, since AFM correlations are decreased due to spin frustration. Our results in Fig. 1 are in agreement with results on a square lattice [13, 26, 28–30] showing (nearly) diverging $\chi_c$ and non-diverging $\chi_c$ on a triangular lattice [31]. Additionally the (spin frustrating) diagonal hopping reduces the divergence of $\chi_c$ and may explain differences in $\chi_c$ between cuprate families [32]. From this picture it is evident that AFM correlations can increase $\chi_c$ at finite dopings, which is discussed also by Imada [33] and indicated in results of Ref. 34. The increase of $\chi_c$ at low $p$ and $T$ due to AFM correlations is supported also by simultaneous decrease of uniform spin susceptibility $\chi_s$ in the same regime (see Figs. 1(c,d) and Supp. [24]). Furthermore, $\chi_c$ may also relate to the pseudogap [24], as is, e.g., discussed also by Sordi et al. [35–37].

In the following we show that strong $p$ and $T$ dependence of $\chi_c$ governs also $\rho = 1/\sigma$, as is expected from Eq. (1), in particular at high $T$ and low $p$. This is the main result of this work. We first calculate via FTLM [13, 25] optical conductivity $\sigma(\omega)$ as a dynamical current-current correlation function [24] and then extract DC conductivity $\sigma = \sigma(\omega \to 0)$. This method gives exact results for large $T > T_{\text{fs}}$, while at low $T < T_{\text{fs}}$ finite size effects appear. We do not show low $T$ results which are potentially affected by finite size [24].

In Fig. 2 we show $T$ dependence of calculated $\rho$, $1/\chi_c$ and $1/D$, which we estimate from $1/D = \rho \chi_c/e^2$. It is clear that the $T$ and $p$ dependence of $\rho$ is dominated by $T$ and $p$ dependence of $\chi_c$, while $D$ shows a rather weak and modest dependence. The most intriguing dependence of $\rho$ and $1/\chi_c$ appears in the low-$T$ and low-$p$ regime, where both quantities from small values at low $T$ show strong and close to linear-in-$T$ increase towards large insulating ($p = 0$) values. This originates in the move of $\mu$ into the charge gap with increasing $T$ to conserve the density of electrons as discussed in Supp. [24]. On the other hand, $1/D$ shows much more mundane dependence with weak increase with increasing $T$ and small decrease with increasing $p$ in the whole shown regime. For example, $1/\chi_c$ increases for $p = 0.05$ by about a factor of 2 from $T = 0.5t$ to $T = t$, while $1/D$ shows a small increase by about 10%. Similarly for $p = 0.05$, $\rho$ shows a weak non-monotonic $T$ dependence, which becomes even more non-monotonic for lower $p$, and originates in non-monotonic $T$ dependence of $1/\chi_c$. It can not be understood via monotonic $1/D$. In the $T \to 0$ limit one expects scattering rate and $1/D$ to go to 0 and therefore to strongly influence $\rho$. We do not reach this regime since finite size effect appear first, but we stress that system size dependence (not shown) of our results suggest weak $T$ dependence of $D$ and dominance of $\rho$ by $\chi_c$ even for lower $T$ than shown in Fig. 2 (see Supp. [24] for more details). At lowest $T$ and dopings ($p < 0.15$), $1/\chi_c$ shows additional suppression which presumably originates in the AFM correlations as discussed in Fig. 1, and could be related to the pseudogap.

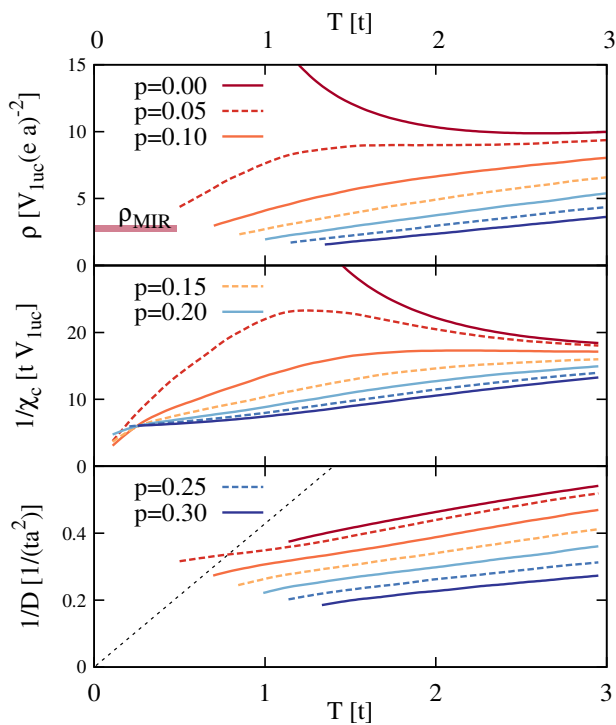

Figure 2. (Color online) Resistivity $\rho$ (upper panel) of doped Mott insulator shows strong $T$ and doping ($p = 1 - n$) dependence at high $T$, and is dominated by the $T$ and $p$ dependence of inverse charge susceptibility $1/\chi_c$ (middle panel). $1/\chi_c$ clearly shows similar behavior as $\rho$. On the other hand, in the same regime diffusion constant ($D$) has a very weak $T$ and $p$ dependence. In particular, see the regime of $p = 0.05$ and $0.5t < T < t$ in which $\rho \propto T$ originating in $1/\chi_c \propto T$ and its increase by about a factor of 2, while $D$ increases only by about 10%. Estimated upper bound on $1/D$ suggested by Hartnoll [18] is shown with black dashed line in the lower panel. Results are calculated for square lattice with $U = 10t$ and regimes potentially influenced by finite size effects are not show.

From $D$ we can estimate the mean free path $l$ via $D = \langle v \rangle l/2$. The estimate of average velocity $\langle v \rangle$ is not straightforward and we use $\langle v \rangle = 8ta/\pi$, which together with the values of $D$ from Fig. 2 gives $l$ of the order of lattice spacing, $l \sim a$. Solely from $l$ one would therefore not expect the crossing of the MIR limit [1] $\rho_{\mathrm{MIR}} = \sqrt{2\pi}V_{1uc}/(e^2 a^2 \sqrt{1-p})$. However, $\rho$ does cross $\rho_{\mathrm{MIR}}$ due to strong increase of $1/\chi_c$ towards the insulating values with increasing $T$ as shown in Fig. 2. The crossing of $\rho_{\mathrm{MIR}}$ without any change or indication has therefore a natural explanation in our picture in terms of decreasing

$\chi_c$. And further, since there in no limit for $1/\chi_c$ close to the insulating regime ($1/\chi_c = \infty$), one does not expect an upper limit for the resistivity.

Hartnoll suggested [18] that the diffusion is bounded with the upper limit for $1/D < k_B T/(\hbar v_F^2)$. Here $v_F$ is the Fermi velocity and by taking its bare band estimate $v_F \sim 8ta/\pi$ the calculated values of $1/D$ shown in Fig. 2 strongly violate the upper bound. However, by using the renormalized value $v_F \sim Z8ta/\pi$, with estimated renormalization $Z = 0.6$ [24] and showing the obtained upper bound with black dashed line in Fig. 2, the violation region moves to lower $T$ and becomes less apparent. Still this indicates possible violation of the diffusivity bound at low $T$, but for a strict test $v_F$ should be calculated separately [24].

Strong $T$ and $p$ dependence of $\chi_c$ shown in Fig. 2 appears for any doped insulator, including Mott insulator, band insulator and semiconductor, as long as $n$ is fixed and not $\mu$. In doped semiconductor the behavior is affected also by, e.g., acceptor levels $\epsilon_a$ above the valence band $\epsilon_v$, which are localized ($D = 0$) [24]. However, the regime of small dopings and $\epsilon_v - \epsilon_a \ll T < E_g$, for which the chemical potential is in the charge gap $E_g$, has a text-book [38] density of conducting valence band holes $p_v = P_v e^{(\epsilon_v - \mu)/T}$. Similar relation between $p$, $T$ and $\mu$ is expected also in doped Mott insulators and has indeed been proposed also for cuprates [26, 39] and readily leads to $1/\chi_c = T/p$. We further suggest this is the origin of $\rho \propto T/p$ at high $T$. We however stress, that our results are only close to this picture since hole doping shifts $\mu$ into the lower band, while in semiconductors it stays above the valence band maximum.

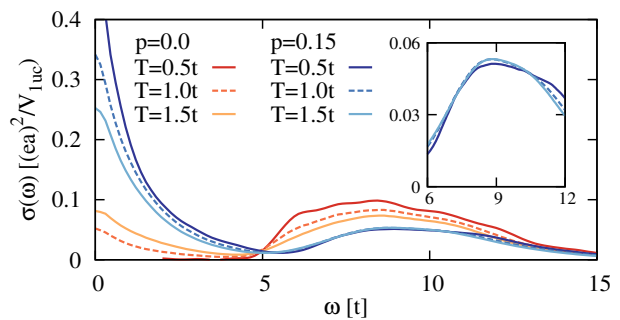

Figure 3. (Color online) Real part of conductivity $\sigma(\omega)$ for two dopings ($p = 0$ and $p = 0.15$) and for several high $T$ ($0.5t$, $t$ and $1.5t$). $p = 0.15$ case shows spectral weight transfer from low $\omega$ to high $\omega$ (see inset) with increasing $T$, and is simultaneously becoming more similar to insulating case. $p = 0$ case shows opposite transfer of weight from high $\omega$ to low $\omega$ (activated behavior) with increasing $T$. Results are for square lattice and $U = 10t$.

Another feature arising naturally in the proposed picture is the optical spectral weight transfer, namely the decrease of $\sigma(\omega)$ at low $\omega$ and increase of $\sigma(\omega)$ at higher $\omega$ with increasing $T$ for finite doping and vice versa for $p = 0$. This is shown in Fig. 3 and is due to finite $p$ behavior becoming more insulating like with increasing $T$. Simultaneously the spectra for $p =$

0.15 and $p = 0$ shown in Fig. 3 are getting similar at low and high $\omega$. We note that some transfer can be suppressed by the decrease of the integrated spectral weight (absolute kinetic energy) with increasing $T$.

*Experiments.* It is of utmost importance to see whether experiments show any confirmation or indication of the proposed picture. We first need to determine the temperature denoted $T_{\text{coh}}$, above which $\chi_c$ has strong $T$ dependence. We estimate [24] $T_{\text{coh}} \sim pW$ where $W$ is half-band width and $T_{\text{coh}} \to 0$ as $p \to 0$. E.g., for $t = 0.35$ eV and $p = 0.1$ we estimate $T_{\text{coh}}$ of the order of 1000 K. However, for diverging $\chi_c$ at the MIT (see Figs. 1 and 2) $T_{\text{coh}}$ can be reduced by an order of magnitude, e.g., to the order of 100 K for the chosen case. Similar estimate is discussed also by Imada [33].

By using photo emission spectroscopy (PES) it is possible to measure $T$ and $p$ dependence of $\mu$ allowing for direct experimental estimate of $\chi_c$. Results for cuprates show strong increase of $\mu$ with increasing $T$ already for $T \sim 200$ K, strong $p$ dependence at small dopings (see Fig. 8.8b in Ref. 40), suggest very large values of $\chi_c$ in the underdoped regime [41, 42] and suggest strong and close to linear-in-$T$ dependence of $1/\chi_c$ [43]. All these results are in qualitative agreement with our results and more importantly suggest that $\rho$ can indeed be influenced by strong variations of $\chi_c$ already at quite low $T$. Unfortunately the direct comparison of experimental $\rho$ and $\chi_c$ as well as extraction of $D$ is unfeasible due to large uncertainties of PES data. Additionally other effects, e.g. Madelung potential [44, 45], can potentially influence the PES data.

Since $\chi_c$ is a basic static quantity, which measures the change of electronic density also due to the external electric potential it influences also many other quantities. For example, Thomas-Fermi screening length of the external potential $l_{TF} \propto 1/\sqrt{\chi_c}$ [38] should have strong $T$ and $p$ dependence and the former is indeed supported by the experiment [46]. It is also expected that $\chi_c$ plays a considerable role for a phonon softening [47] and even for the superconductivity [30, 33].

On the other hand, one can indirectly compare at least on qualitative level the experimental high-$T$ $\rho$ and our $1/\chi_c$. Data on various cuprates [1, 48–52] show $\rho \sim aT$ at high $T$ with the prefactor $a$ increasing with decreasing $p$ in good qualitative agrement with our results in Fig. 2. Similar increase of linear-in-$T$ part of $\rho$ has been observed also in the low-$T$ regime [4, 53, 54], but we stress that at such low $T$ $1/\tau$ and $D$ are expected to give the dominant $T$ dependence.

Behavior of $\rho$ and $\chi_c$ shown in Fig. 2 is not expected just in above discussed Mott insulators, but also for doped band insulators. Remarkably similar $p$ and $T$ dependence has been lately reported for the electrostatically doped band insulator $MoS_2$ [55]. Furthermore, in nickelates the bad-metallic behavior is related [10] to decreasing effective carrier density $n_{\text{eff}}$ with increasing $T$ while measured $\tau$ and effective mass showed rather weak $T$ dependence. This is closely related to our decreasing-in-$T$ $\chi_c$ and weak $T$ dependence of $D$. Also the spectral weight transfer discussed above (see Fig. 3) is in qualitative agreement with optical conductivity for cuprates and many other bad-metallic materials [52, 56–58].

*Conclusions.* We propose an explanation of large and linear-in-$T$ resistivity of doped insulator in the bad-metallic regime in terms of much simpler static quantity, namely charge susceptibility $\chi_c$. Furthermore, since $\chi_c$ should have similar $T$ and doping dependence for any charge insulator (with exception of insulators due to spatial localization of electrons) similar features in resistivity are expected for Mott and band insulators, semiconductors as well as AFM slater insulators. We further discuss that in this simple picture the coefficient of linear-in-$T$ resistivity is close to being proportional to $1/p$, while the diffusion constant $D$ shows in the corresponding regime weak doping and $T$ dependence. Same holds for the mean free path, which we estimate to a few lattice spacing. Crossing of MIR limit observed in resistivity can be naturally understood as $\chi_c$ becoming small with increasing $T$ due to proximity of insulating state. We note that our picture with weak $T$ dependence of $D$ is opposite to the main suggestion that $D \propto 1/T$ by Hartnoll [18] and is closer to the results by Pakhira and McKenzie [23], where $D$ and its possible lower bound were discussed for a half-filled case [24] within a dynamical-mean field (DMFT) approach. Our data suggest a possible violation of the diffusivity bound [18], but a strict test would require a separate calculation of the Fermi velocity [24] which is beyond the scope of this work. Recently, importance of $\chi_c$ for $\rho$ was supported also with a high-$T$ series expansion approach [24, 59]. We further discuss how AFM correlations increase $\chi_c$ at low $T$, which might lead to pseudogap features in the resistivity.

*Future challenges.* It remains theoretical and experimental challenge to establish in which $T$ and doping regime the resistivity is dominated by $\chi_c$ and in which by $D$. Theoretical challenge is to reach lower $T$ and in this respect continuous improvements of the insight into the phenomena, of analytical and numerical techniques as well as computational power is promising. We further comment in the Supp. [24] previous works [60–63], which indicate the importance of static effects.

Even more important is to experimentally establish the behavior of $\chi_c$ and $D$ with $T$ and $p$. In this respect PES, Thomas-Fermi screening length and other possibilities are invaluable. $D$ has been already measured in cuprates [64], but unfortunately for nonequilibrium quasi-particles. This makes it inapplicable to our analysis. Measurements of $D$ are very challenging but also highly desirable. Experimental deconstruction of $\rho$ onto $\chi_c$ and $D$ would further pose strong constraints on the theories and would differentiate between them. They would for example offer insight on the questions like; is there a quantum critical point [15, 16], is diffusion bounded as motivated by holographic duality [18], is there a regime with universal behavior of diffusion constant [65], is the pseudogap a new phase [66], is the nature of the insulator Mott, band or (AFM) slater like. Deconstruction of $\rho$ onto $\chi_c$ and $D$ therefore offers an exciting, promising and yet to be fully explored approach towards better understanding of discussed phenomena, both from theoretical and experimental point of view.

I acknowledge helpful discussions with Ross McKenzie,

Peter Prelovšek, Takami Tohyama, Jakša Vučičević, Veljko Zlatić, Jernej Mravlje, Masatoshi Imada, Neven Barišić, Giovanni Sordi, Ivan Božović and Joseph Orenstein. This work was supported by Slovenian Research Agency Grant No. Z1-5442 and Program P1-0044.

# Supplemental Material for "Bad-Metallic Behavior of Doped Mott Insulators"

Jure Kokalj

*J. Stefan Institute, SI-1000 Ljubljana, Slovenia and*
*Faculty of Civil and Geodetic Engineering, University of Ljubljana, SI-1000 Ljubljana, Slovenia*

(Dated: December 28, 2016)

## I. COMMENTS ON NERNST-EINSTEIN RELATION

The generalized dynamical charge susceptibility can be written as [S1–S3]

$$\chi(q,\omega) = \chi_q - \frac{\omega \chi_q}{\omega + iq^2 D(q,\omega)}. \quad \text{(S1)}$$

Here $\chi_q$ is a static charge susceptibility and $D(q,\omega)$ is generalized (to finite $q$ and $\omega$) diffusion constant. This form of $\chi(q,\omega)$ has the right long wave length limit $q \to 0$ with $\chi'(q \to 0, \omega = 0) = \chi_c$ being the static uniform charge susceptibility and $\chi''(q \to 0, \omega)/\omega = \chi_c \delta(\omega)$ reflecting particle number conservation. It describes the diffusive behaviour with diffusive broadening of $\delta(\omega)$ peak in $\chi''(q \to 0, \omega)/\omega$ with increasing $q$ from $q=0$. Eq. (S1) can be taken also as the definition of generalized diffusion constant $D(q,\omega)$. Using such form for $\chi(q,\omega)$ one obtains Nernst-Einstein relation by simply using the well known relation to the conductivity

$$\sigma = e^2 \lim_{\omega \to 0} \lim_{q \to 0} \omega \chi''(q,\omega)/q^2 = e^2 \chi_c D. \quad \text{(S2)}$$

Here we denoted the standard diffusion constant with $D = D(q=0, \omega=0)$. We note that the above considerations are valid regardless of the phase realized in the system as long as the behavior is diffusive, e.g., for Fermi liquids with coherent quasiparticles and for bad metals with incoherent transport.

However, there is still some chance that the behavior is non-diffusive. With such a possibility in mind, we take the Nernst-Einstein relation as a postulate, which allows us to discuss the bad-metallic behavior in terms of diffusion constant and charge susceptibility as well as to discuss the suggested [S4] bound on diffusion constant.

## II. METHOD AND FINITE SIZE EFFECT

We use the finite temperature Lanczos method (FTLM) [S5, S6], which is based on the diagonalization of small clusters. Within the FTLM we calculate the optical conductivity with a general form

$$\sigma(\omega) = 2\pi e^2 D_c \delta(\omega) + \sigma_{\text{reg}}(\omega), \quad \text{(S3)}$$

by evaluating the regular part as the dynamical current-current correlation function

$$\sigma_{\text{reg}}(\omega) = \frac{e^2(1 - e^{-\beta\omega})}{N\omega} \text{Re} \int_0^\infty dt e^{i\omega t} \langle j_\alpha(t) j_\alpha(0) \rangle, \quad \text{(S4)}$$

where the current operator is $j_\alpha = i \sum_{\langle i,j \rangle, s} t(R_{i,\alpha} - R_{j,\alpha}) c^\dagger_{j,s} c_{i,s}$, with $\alpha$ being either $x$ or $y$ and $R_{i,\alpha}$ the coordinate of site $i$. The charge stiffness $D_c$ is evaluated by using the optical sum rule $\int \sigma(\omega) d\omega = \pi e^2 \langle \tau_{\alpha\alpha} \rangle / N$ with $\tau_{\alpha\alpha} = \sum_{\langle i,j \rangle, s} t(R_{i,\alpha} - R_{j,\alpha})^2 c^\dagger_{j,s} c_{i,s}$.

By using FTLM we are limited by finite size effect at low $T$. Even though we apply techniques like averaging over twisted boundary conditions to reduce finite size effects, some still remain at low $T < T_{\text{fs}}$. In $\sigma(\omega)$ finite size effects appear in the form of finite charge stiffness $D_c$ at finite $T$, while we know $D_c$ should be zero due to non-integrability of the model. $D_c$ becomes smaller with cluster size but at low $T$ ($< T_{\text{fs}}$) the associated delta peak, $2\pi e^2 D_c \delta(\omega)$, represents considerable part of the total spectral weight. Due to some numerical uncertainty, we judge that the finite size effects in $\sigma(\omega)$ are negligible and $D_c = 0$, if the relative weight of $D_c$ peak to the total spectral weight is less than 1% [S5].

In Fig. 2 in the main text we show that $\rho$ is dominated by $1/\chi_c$ and that $D$ shows weak $T$ dependence. We show result only for a certain range where $T > T_{\text{fs}}$. We further argue that the weak $T$ dependence of $D$ extends to even lower $T$, if one considers system size dependence. This comes from the decrease of $D_c$ with increasing system size (our criteria for finite size effect) in a manner that does not alter $\sigma(\omega)$ at low $\omega$ much, and for this low $\omega$ the almost $T$ independent behavior of $D$ shown in Fig. 2 in the main text extends to lower $T$, which we, e.g., estimate to $T \sim 0.2t$ for $p = 0.05$. On the other hand, finite size effects for $\chi_c$ are smaller, as indicated by showing $1/\chi_c$ to much lower $T$ in Fig. 2 in the main text.

## III. AVERAGE $\chi_C$ FOR FINITE DOPINGS AND RENORMALIZATION $Z$.

Average $\bar{\chi}_c$ for finite dopings ($n \neq 1$) can be in the case of a finite charge gap $\Delta_c$ calculated as $\bar{\chi}_c = 2/(\mu_{\text{max}} - \mu_{\text{min}} - \Delta_c)$. This takes into account that the total change of density of electrons from completely empty to completely filled band is 2 and that the density does not change for $\mu$ within the charge gap $\Delta_c$. If $U$ is increased to $U + \delta U$ deep in the Mott insulating regime $U \ll t$, then $\mu_{\text{max}} \to \mu_{\text{max}} + \delta U$ and $\Delta_c \to \Delta_c + \delta U$ leading to $\bar{\chi}_c \to \bar{\chi}_c$. Therefore, with increasing $U$ the averaged $\bar{\chi}_c$ is not changed, but we note, that details of $\chi_c$, or $\chi_c$ for a particular doping, can still change with $U$.

For the estimate of the upper bound on $1/D < k_B T/(\hbar v_F^2)$ suggested by Hartnoll [S4], we need to estimate the value of $v_F$. Taking the estimate for non-normalized value $v_F \sim 8ta/\pi$ our results for $1/D$ strongly violate the Hartnoll's bound. Taking instead the normalized $v_F \sim Z8ta/\pi$ could make the bound valid. We therefore need to estimate the normalization $Z$. It is related to the self-energy, and since we have not calculated it, we will use an approximation $\chi_c = Z\chi_c^0$ [S6], with $\chi_c^0$ being a bare band noninteracting charge susceptibility. This is a crude approximation neglecting $\mu$ and wave-vector ($k$) dependence of self-energy. Estimating $\chi_c = 2/(2W + U - \Delta_c)$, with the half-band width $W = 4t$, $U = 10t$ and $\Delta_c = 4t$ from Fig. 1a in the main text, and $\chi_c^0 = 1/W$ we obtain an estimate $Z \sim 0.6$. For such value the results for $1/D$ show slight violation of the diffusivity bound (see Fig. 2 in the main text). In a similar way one can estimate $Z$ via $\chi_c$ even for each $T$ and doping $p$ separately. With such estimate our data do not show any violation of the diffusivity bound, but we stress that such estimate is problematic, since $\chi_c$ with increasing $T$ approaches gapped behavior leading to decreasing $Z$. This is the density of states effect, while one expects increasing $Z$ with increasing $T$ from the behavior of self energy. Further can the renormalized $v_F$ stay rather constant with decreasing doping $p$ at least in nodal direction [S7] while such estimate gives substantial dependence. Therefore, for a strict test of the diffusion bond, $v_F$ should be calculated separately, e.g., by obtaining renormalization via self energy calculation.

## IV. ORIGIN OF STRONG REDUCTION OF CHARGE SUSCEPTIBILITY WITH INCREASING TEMPERATURE

The behavior of $\chi_c$ shown in Fig. 1c in the main text and the move of $\mu$ for small $p$ into the gapped regime with increasing $T$ originates in the following. For small dopings the increase of $T$ populates states according to the Boltzmann statistics and the states at half-filling and smaller density ($n \leq 1$) are relatively low in energy considering $\mu$ for chosen $p$ and can therefore be easily populated, while states with $n > 1$ are higher in energy for about a charge gap $\Delta_c$ and cannot be efficiently populated for $T < \Delta_c$. Therefore the only way for the system to conserve the number of electrons with increasing $T$ is to move $\mu$ into the gap and increasingly populate the half-filled ($n = 1$) states. From this, one can expect that in this $T$ regime the behavior of the doped system will approach the behavior of the gapped one with increasing $T$. We stress that this situation appears for any doped charge-gapped system, as long as the density of electrons is conserved (e.g. semiconductor, band insulator, Mott insulator, etc.).

## V. SEMICONDUCTING PICTURE

In doped semiconductor $\chi_c$ shows two activated insulating regimes [S8], e.g., at low $T$ due to $\mu$ being in the gap between valence band $\epsilon_v$ and acceptor levels $\epsilon_a$, while at higher $T$, due to $\mu$ being close to the middle of the band gap $E_g$. Similar behavior is observed also in $\rho$, but for proper behavior also $D$ should be considered, in particular, since localized doping levels contribute to $\chi_c$ but have $D = 0$. As already mentioned in the main text, the regime of small doping and $\epsilon_v - \epsilon_a \ll T < E_g$ for which the chemical potential is in the charge gap $E_g$ has density of conducting valence band holes given by $p_v = P_v e^{(\epsilon_v - \mu)/T}$ [S8]. This readily leads to $\chi_c = p_v/T$ and dictates $\rho \propto T/p_v$. In semiconductor such behavior of $\rho$ can be mask by much stronger delocalizing activated (exponential) behavior of localized states, but similar relation for the density of conducting holes can be expected also in doped Mott insulators. There the doped charges are itinerant and therefore strongly decrease $\rho$ at lowest $T$ being in contrast to semiconducting behavior.

## VI. CHARGE AND SPIN SUSCEPTIBILITIES AT LOW DOPING

In Fig. 1d in the main text we show spin susceptibility $\chi_s$ for several $p$ and $T$. In the low-$p$ regime ($p = 0.5 - 0.10$) and low $T$ ($< 0.3t$), $\chi_s$ shows a suppression, which is known [S9] to be due to the increased antiferromagnetic correlations or a pseudogap effect. In the same regime charge susceptibility $\chi_c$ shows an increase (see Fig. 1c in the main text). To make this correlation between decreased $\chi_s$ and increased $\chi_c$ in the underdoped regime more explicit we show in Fig. S1 the sum of both susceptibilities, $\chi_c + \chi_s$. From the sum it is seen that the sum behaves smoothly and very similarly at low $T$ for all low dopings $p$, which suggest that for $\chi_c + \chi_s$ the suppression of $\chi_s$ is compensated with increase of $\chi_c$ and that these two susceptibilities are correlated. Therefore, it seems that the increased antiferromagnetic correlations decrease $\chi_s$ and at the same time increase $\chi_c$.

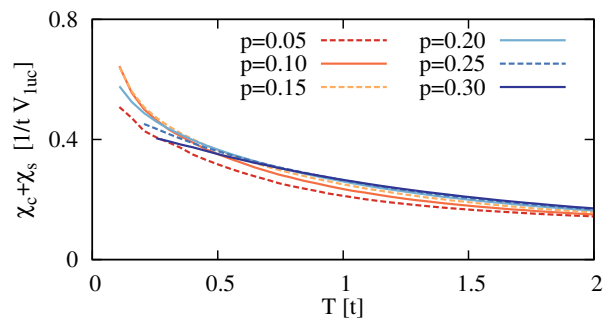

Figure S1. Sum of susceptibilities, $\chi_c + \chi_s$, shows similar behavior in the low $T < 0.3t$ regime for all low dopings $p < 0.15$. This exhibits correlation between suppression of $\chi_s$ in this regime with increase of $\chi_c$ in the same regime (see Fig. 1c and 1d in the main text). Calculated for square lattice and for $U = 10t$.

## VII. ESTIMATE OF $T_{\text{coh}}$

Here we discuss in more detail the temperature, which we denote by $T_{\text{coh}}$, above which $\chi_c$ has strong $T$ dependence.

Estimating $T_{\text{coh}} \sim |\mu(p) - \mu_c|_{T=0}$ with $\mu(p)$ and $\mu_c$ being the $T = 0$ chemical potential at doping $p$ and critical chemical potential for the transition to the insulating state, respectively, we obtain for non-diverging $\chi_c$ that $T_{\text{coh}} \sim pW$. $W$ is the half-band width. $T_{\text{coh}}$ therefore goes to 0 as $p \to 0$ and can be much lower than $W$, e.g., for $t = 0.35$ eV and $p = 0.1$ we estimate $T_{\text{coh}}$ of the order of 1000 K. However, for diverging $\chi_c$ at the MIT, $T_{\text{coh}}$ is reduced by an order of magnitude. For the same case ($t = 0.35$ eV and $p = 0.1$) it is reduced to the order of 100 K. Similar picture and estimate $T_{\text{coh}} \sim t^* p^2$ (with $t^*$ being renormalized hoping) is discussed also by Imada [S10], who also estimates $T_{\text{coh}}$ to around 100 K for $p = 0.1$.

## VIII. HALF-FILLED CASE

For a half-filled case one expects bad-metallic behavior for $U$ close to the critical $U_c$ at which the Mott metal-insulator transition appears and is, e.g., realized in organic charge transfer salts [S11]. Previous Hubbard model results for $\chi_c$ [S6, S12] show weak $T$ dependence at high $T$ and for $U < U_c$, which differs from strong $T$ dependence of doped Mott insulators discussed in the main text. Therefore one could expect that in the half-filled case $\chi_c$ plays a smaller role. One should however be careful. The theoretical $T$ dependence in Ref. S6 is limited due to finite size effects to finite $T$, while at lowest $T$, $\chi_c$ could close to $U_c$ have stronger $T$ dependence. This is indeed indicated in DMFT results (see Fig. 3 in Ref. S12). On the insulating side ($U > U_c$) $\chi_c$ has strong activated $T$ dependence and similar behavior is expected for resistivity, in particular if the charge gap is small. On the other hand, the high-$T$ experimental resistivity (see, e.g., inset in Fig. 2 in Ref. S13 and Fig. 6 in Ref. S14), similarly as theoretical $\chi_c$, shows saturating or weak $T$ dependence. $\chi_c$ might offer an explanation of the violation of the Mott-Ioffe-Regel limit. Furthermore, the experimental resistivity strongly varies with approaching to, or departing from, the metal-insulator 1st order transition line or the widom line in $U$-$T$ or ($P$-$T$) phase diagram (see, e.g., Ref. S15) and it is natural to expect strong corresponding changes in $\chi_c$ as well. Proper determination of the relative importance of $D$ and $\chi_c$ for half-filled case is therefore and open future challenge.

## IX. COMMENT ON OTHER WORKS INDICATING IMPORTANCE OF STATIC EFFECTS

Importance of the static effects in the bad-metallic regime has been already advocated, e.g., in terms of quasiparticle renormalization [S16] or carrier density [S17], but correspondence with our picture needs to be further clarified. Our picture is also close to the one presented by Boyd et al. [S18], who discusses $\rho$ in terms of transport relaxation time, which is presumably dominated by the effective density of states, but distinct behavior of static and dynamic quantities has not been discussed. Indication of a weak $T$ dependence of $D$ at higher temperatures comes also from a DMFT study [S19] where a weak $T$ dependence of the imaginary part of self-energy was observed.

In Ref. S20 a doped Hubbard model is considered in a limit of infinite $U$ (or $J = 0$) with a high-$T$ series expansion. Therefore the activated behavior for the half-filled case is not captured, including the low-doping non-monotonic $T$ dependence of $\chi_c$ and $\rho$. Due to vanishing of $J$, also the effects of AFM correlations are missing, with the most prominent strong decrease of $1/\chi_c$ at low $T$ in second panel of Fig. 2 in the main text. Similar decrease, due to AFM correlations, is expected also in the resistivity. On the other hand, one expects that the regime of Ref. S20 would correspond in our case to the regime of $T > J$, but below the charge gap $T < \Delta_c$, where indeed similar conclusions on $\rho \propto 1/\chi_c \sim T/p$ were obtained.

---